\documentstyle[aps,pra,epsfig,eqsecnum,twocolumn]{revtex}
\newcommand{\be}{\begin{equation}}
\newcommand{\ee}{\end{equation}}
\newcommand{\bea}{\begin{eqnarray}}
\newcommand{\eea}{\end{eqnarray}}

\begin{document}
\draft

\wideabs{
\title{Collisionless collective modes of fermions in magnetic traps}

\author{P. Capuzzi and E. S. Hern\'andez$^1$ }

\address{Departamento de F\'{\i}sica, Facultad de Ciencias
Exactas y Naturales,
Universidad de Buenos Aires, RA-1428 Buenos Aires, \\
$^1$ and Consejo Nacional de Investigaciones Cient\'{\i}ficas y
T\'ecnicas, Argentina}

\date{\today}
\maketitle
\begin{abstract}

We present a Random-Phase-Approximation formalism for the collective spectrum
of two hyperfine species of dilute $^{40}$K atoms, magnetically trapped at
zero temperature and
subjected to a  repulsive $s$-wave interaction between atoms with different
spin projections.  We examine the density-like and the spin-like oscillation
spectra, as well as the transition density profiles created by external
multipolar fields. The zero sound spectrum is always fragmented and
the density and spin channels become clearly distinguishable if the trapping
potentials acting on the species are identical. Although this distinction is
lost when these confining fields are different, at selected excitation
frequencies the transition densities may display the signature of the channel.

\end{abstract}
}
\narrowtext
\section{Introduction}

The realization of Bose-Einstein condensation of alkali atoms in magnetic
traps triggered substantial experimental efforts, aimed at  cooling fermion
isotopes below their Fermi temperatures. In particular, lithium and potassium
have been trapped and cooled \cite{potasio,litio,pwave},  and a recent
experiment displays unambiguous   evidence of quantum degeneracy effects at
temperatures around half the Fermi temperature of  $^{40}$K
atoms\cite{demarco} in a harmonic trap. On the other hand, various
theoretical descriptions of thermodynamical properties of confined, cool free
fermions have been presented, either in the semiclassical limit\cite{butts} or
with explicit consideration of quantum shell effects \cite{schneider}.  In
Ref. \cite{bruun1} a Hartree-Fock (HF) calculation of the mean field spectrum
of two hyperfine fermionic species subjected to $s$-wave attraction has been
forwarded.  Moreover, in view of the presence of both fermion and boson
isotopes in natural alkali samples, the consequences of their coexistence and
mutual coupling at zero temperature in the magnetic trap is a topic of current
interest\cite{molmer,amoruso1,vichi1,vichi2,viverit,minguzzi}, as well as the
possible occurrence of BCS-like superfluid states driven by attractive
interactions\cite{stoof,houbiers,baranov1,bruun2,baranov2,zhang}.

	An important step towards a full understanding of the behavior of
coexisting hyperfine species is the knowledge of their collective excitation
spectrum. In this context, the collisionless modes of an extended system  with
various hyperfine levels have been examined in the frame of Landau's theory of
Fermi liquids\cite{yip} and the zero-sound collective spectra of two species
of confined fermions have been computed resorting to sum rules\cite{vichi3}.
Assuming local equilibrium of a Fermi gas, described by a Thomas-Fermi (TF)
approximation, the linearized hydrodynamic equations can be analytically
solved both in the degenerate and in the classical limits\cite{bruun3}. A
related hydrodynamic-like approach based on the equations of motion for the
first and second moments of the fermionic Wigner distribution allowed to
compute the oscillation modes of an  isotope with one \cite{amoruso2} and two
spin components \cite{amoruso3}.

	Since the two hyperfine species of trapped fermionic atoms are very
dilute, it is not clear that  when the system is excited by a low frequency
external field, the $s$-wave interaction --which is supposed to play a
relevant role in thermalization during the evaporative cooling process-- may
permit the trapped gas to equilibrate locally and develop hydrodynamic
oscillation modes.  In particular, it has been shown that for $^{40}$K, at
least $10^8$ atoms should be cooled in each hyperfine state in order to reach
the hydrodynamic regime in the degenerate quantum phase. It is then important
to focus upon the  study of collisionless excitation spectra of these systems,
seeking to improve the understanding of their low temperature behavior, as
well as the evolution between the collisionless and the hydrodynamic regimes,
as increasing temperature suppresses Pauli blocking effects and enhances the
collision rate.  For this sake, in this work we derive a
Random-Phase-Approximation (RPA) description of the collective modes of two
species of fermions in a harmonic well with mutual $s$-wave coupling at zero
temperature,  and apply the formalism to the computation of density-like and
spin-like fluctuations. The paper is organized as follows: the specific  RPA
frame and the extraction of the elementary excitation spectrum of
quasiparticles in a mean field approach are discussed in Sec.~\ref{sec:rpa} .
In Section \ref{sec:model_zero} we propose a simple model to describe the zero
sound modes that permits to analyze various major features of the expected RPA
spectra. Typical calculations of collective spectra for the lowest
multipolarities are presented and discussed in Sec. \ref{sec:calcula}, while
Sec. \ref{sec:resumen}  summarizes our main conclusions.

\section{The Random-Phase-Approximation for a trapped fermion system}
\label{sec:rpa}

	We assume that the trapped atom system consists of noninteracting
quasiparticles (qp's) in a mean field. Throughout this paper this is referred
to as free system, which can be excited by an external field so
that particle-hole (ph) pairs involving, in principle, both hyperfine species
($\sigma_1, \sigma_2$), are created with energy $\Omega$. The spectral
properties of this nonhomogeneous free system are contained in the free ph propagator
$G_0^{\sigma' \sigma^{-1}}(\Omega)$, where the labels $\sigma$ and $\sigma'$
stand for either $\sigma_1$ or $\sigma_2$.   

We also suppose that a ph effective interaction $V_{\rm ph}^{\sigma
\sigma^{-1} \sigma' \sigma'^{-1}}$ acting between the qp's gives rise to the
dressed propagator for ph pairs according to the RPA integral equation
\cite{fetter}
\bea
G^{\sigma' \sigma^{-1}}(\Omega) &=& G_0^{\sigma' \sigma^{-1}}(\Omega)
\nonumber
\\
& +& \sum_{\tau \tau'} G_0^{\sigma' \tau^{-1}}(\Omega)\,V_{ph}^{\tau 
\tau^{-1} \tau' \tau'^{-1}}\,G^{\tau' \sigma^{-1}}(\Omega)
\label{rpa}
\eea

It is worthwhile noticing that the system under consideration is both
nonhomogeneous \cite{bertsch,Casas-Stringari,CH} and polarized \cite{HNP,GHN},
so that the present development merges the corresponding formalisms as shown
below.  Hereafter, we consider longitudinal excitations 
involving propagation of ph pairs of the same spin kind, created by
spin-symmetric ($s$) and spin-antisymmetric ($a$) multipolar operators of the form 
\be
O^{\dagger}_{s,a} = \sum_{i=1}^{N_1}\,O^{\dagger}_i \pm
\sum_{i=1}^{N_2}\,O^{\dagger}_i
\label{eq:operators}
\ee
with 
\be
O^{\dagger}_i=\left\{
\begin{array}{cr}
r_i^l\,Y_{LM}(\theta_i,\phi_i)& L \neq 0 \\
r_i^2 & L = 0 
\end{array}\right.
\ee
being $N_1$ (resp. $N_2$) the number of trapped atoms of  species $\sigma_1$
(resp. $\sigma_2$), and $L$ the multipolarity of the perturbation.  Notice that a
particle-particle interaction $V({\mathbf r}-{\mathbf r}')$, with ${\mathbf
r}, {\mathbf r}'$  respectively denoting particles with spin projections
$\sigma$ and $\sigma'$, gives rise to a ph  interaction that scatters a
($\sigma \sigma^{-1}$) pair at position ${\bf r}$ into a  ($\sigma'
\sigma'^{-1}$) pair at  ${\bf r}'$, and that only collisions among different
species are allowed . The free ph propagators involved
in longitudinal density fluctuations are diagonal in spin space and thus Eq.
(\ref{rpa})  splits into two equivalent systems of two equations each,
 intrinsic to polarized systems \cite{HNP,GHN}, which in coordinate representation read 
\bea
&&G^{\sigma \sigma^{-1}}({\bf r},{\bf r'})=G_0^{\sigma \sigma^{-1}}({\bf
r},{\bf r'}) \nonumber \\
&&\, +\int d^3r_1 d^3r_2\,
G_0^{\sigma \sigma'^{-1}}({\bf r},{\bf r_1})V_{\rm ph}^{\sigma \sigma'}({\bf
r_1},{\bf r_2}) G^{\sigma' \sigma^{-1}}({\bf r_2},{\bf r'})
\nonumber
\\
&&G^{\sigma' \sigma^{-1}}({\bf r},{\bf r'})= 
\nonumber \\
&&\int d^3r_1 d^3r_2 \,G_0^{\sigma'
\sigma'^{-1}}({\bf r},{\bf r_1})\,V_{\rm ph}^{\sigma' 
\sigma}({\bf r_1},{\bf r_2})\,G^{\sigma \sigma^{-1}}({\bf r_2},{\bf r'})
\nonumber \\ \label{sistema}
\eea
In spatially inhomogeneous systems, it is convenient  to expand both free and 
dressed propagators in multipolar decompositions \cite{bertsch,Casas-Stringari}
\be
G^{\sigma' \sigma^{-1}}({\bf r}, {\bf r}', \Omega)= \sum_L\,G_L^{\sigma' 
\sigma^{-1}}(r, r',\Omega)\,P_L({\hat r} \cdot {\hat r}')
\ee
with the Legendre polynomials $P_L(x)$. The free ph propagator reads 
\bea
&&G_0^{\sigma \sigma^{-1}}({\bf r},{\bf r}', \Omega) = 
\nonumber
\\
&&\sum_{\nu
 \nu'}\,\phi_{\nu}^{\sigma}({\bf r})\,
\left[\phi_{\nu'}^{\sigma}({\bf r})\right]^*\,
\bigl[\phi_{\nu}^{\sigma}({\bf r}')\bigl]^*\,\phi_{\nu'}^{\sigma}({\bf 
r}')\,\chi_{\nu \nu'}^{\sigma}(\Omega)
\eea
with $\phi_{\nu}^{\sigma}({\bf r})$ a single-particle (sp) wave function for 
energy eigenvalue $\varepsilon_{\nu}^{\sigma}$ and $\chi_{\nu \nu'}^{\sigma}$
the generalized susceptibilities in terms of the Fermi-Dirac 
occupation numbers $n(\varepsilon)$
\be
\chi_{\nu \nu'}^{\sigma} = 
\frac{n(\varepsilon_{\nu}^{\sigma})-n(\varepsilon_{\nu'}^{\sigma})}{\Omega - 
\left(\varepsilon_{\nu'}^{\sigma}-\varepsilon_{\nu}^{\sigma}\right) + i\,\eta}
\label{chinunu'}
\ee
Here the label $\nu$ stands for the spherical quantum numbers ($n l m $).
The expressions for the multipole components $G_{0\,L}^{\sigma 
\sigma^{-1}}(r, r', \Omega)$ are given in the Appendix.

Since   (\ref{sistema}) is a matrix equation calling for discretization  in 
radial coordinates, it is convenient  
to map it onto a vector system for the transition densities defined as
\bea
\delta \rho_{LM}^{\sigma' \sigma^{-1}}({\bf r}, \Omega) &=&
\int d {\bf 
r}'\,G^{\sigma' \sigma^{-1}}({\bf r}, {\bf r}', \Omega)\,r'^L\,Y_{LM}({\hat r}')
\nonumber
\\
&=& \frac{4\pi}{2 L+1}\,\delta \rho_{L}^{\sigma' \sigma^{-1}}(r, 
\Omega)\,Y_{LM}({\hat r})
\eea
where
\be
\delta \rho_{LM}^{\sigma' \sigma^{-1}}(r, \Omega) =
\int d r'\,r'^{\,2+L}\,G_L^{\sigma' \sigma^{-1}}(r, r', \Omega)
\ee
The multipolar susceptibility can then be computed as
\bea
\chi_{LM}^{\sigma' \sigma^{-1}}(\Omega) &= &\int d {\bf r}\, 
r^L\,Y^*_{LM}({\hat r})\,
\delta \rho_{LM}^{\sigma' \sigma^{-1}}({\bf r}, \Omega)
\nonumber
\\
&=& \frac{4\pi}{2 L+1}\int  d r\,r^{2+L}\,\delta \rho_L^{\sigma' 
\sigma^{-1}}(r, \Omega)
\nonumber
\\
&\equiv& \chi_L^{\sigma' \sigma^{-1}}(\Omega)
\label{chilm}
\eea

For dilute trapped systems at low temperature we can reasonably represent the
interaction potential by a contact interaction  of the form $g \delta({\bf
r}-{\bf r}')$, being $g=4\pi\hbar^2\,a/m$ with $m$ the mass and $a$ the
$s$-wave scattering length of the interacting atoms. Thus, we obtain  from
(\ref{sistema})

\bea
&&\delta \rho_{L}^{\sigma \sigma^{-1}}(r, \Omega) =
\delta \rho_{0L}^{\sigma \sigma^{-1}}(r, \Omega)
 \nonumber \\
&& + \frac{4\pi\,g}{2 L+1}\int d r'\,r'^2\,G_{0L}^{\sigma \sigma^{-1}}(r, r', 
\Omega)\,
\delta \rho_{L}^{\sigma' \sigma^{-1}}(r', \Omega)
\nonumber 
\\
&&\delta \rho_{L}^{\sigma' \sigma^{-1}}(r, \Omega) = \nonumber \\
&&\frac{4\pi\,g}{2 L+1}\int d r'\,r'^2\,G_{0L}^{\sigma' \sigma'^{-1}}(r, r', 
\Omega)\,
\delta \rho_{L}^{\sigma \sigma^{-1}}(r', \Omega)
\label{dro}
\eea

 Moreover,  in view of (\ref{eq:operators}) we shall consider the symmetric
and antisymmetric density fluctuations for each atom species, 
\be
\delta\rho_{\sigma}^{(s,a)}={\delta\rho^{\sigma\sigma^{-1}}\pm 
\delta\rho^{\sigma\sigma'^{-1}}\over2} 
\ee

This representation permits to distinguish the hereafter called density-like
(symmetric) and spin-like (antisymmetric) fluctuations \cite{pines} as they
usually appear in Fermi liquids.  In fact, a  multipolar operator
$O^\dagger_{s}$ will likely generate a total density-like fluctuation
proportional to $\delta\rho^{s}_{\sigma}+\delta\rho^{s}_{\sigma'}$ while
out-of-phase perturbations produced  by $O^{\dagger}_{a}$ will induce a
spin-like fluctuation proportional to $\delta\rho^{a}_{\sigma} +
\delta\rho^{a}_{\sigma'}$.  In addition, this distinction enables one to
analyze the influence of a given fluctuation on its own propagation, as well
as on the oscillations in the other spin species. 

The numerical procedure consists of solving the
discretized equations (\ref{dro}) by matrix inversion, computing the
susceptibilities (\ref{chilm})  and constructing the total dynamic structure
factors $S^{(s, a)}=-{\rm Im}  \chi^{(s, a)}/\pi$ in both spin
channels\cite{HNP,GHN}, where

\be
\chi_{L}^{(s,a)}  =  \frac{1}{4} \left(\chi_L^{\sigma \sigma^{-1}} + 
\chi_L^{\sigma' \sigma'^{-1}}\pm\chi_L^{\sigma \sigma'^{-1}}
\pm \chi_L^{\sigma' \sigma^{-1}}\right)
\label{chi_s_a}
\ee
The collective spectrum of density-like and spin-like modes 
for a given multipolarity $L$ is indicated by  the poles of
the real part of these responses, or corresponding peaks in the dynamical 
structure factors. 

It is important to remark that this is a very general RPA description of
collective excitations, valid for any system identified by its elementary
excitations with sp  spectrum $\varepsilon_{\nu}$ and states $\phi_{\nu}$, and
by an effective ph interaction $V_{\rm ph}$.  In most applications to quantum
liquids (see for example Refs.\cite{Casas-Stringari,CH,HNP,GHN} and therein),
one starts from a HF   sp eigenspectrum and chooses the ph coupling as the
double functional derivative of the total energy with respect to the sp
density.  The HF spectrum of two hyperfine species of trapped fermions has
been previously investigated in Ref.  \cite{bruun1} for the case of an
attractive coupling between the species, and in the present work we adopt the
same philosophy for a repulsive interaction of strength $g$. The HF spectrum
arises from  the solution of the coupled nonlinear system in spherical
coordinates
\bea
&&\left\{- \frac {\hbar^2}{2 m}\,\frac{\partial^2}{\partial r^2} + 
\frac{\hbar^2\,l (l+1)}{2 m r^2} + \frac{m \omega_{\sigma}^2\,r^2}{2}+g 
\,\rho_{\sigma'}(r)\right\}
\, u_{nl}^{\sigma}(r)
\nonumber
\\
&&= \varepsilon_{nl}^{\sigma}\,u_{nl}^{\sigma}(r)
\label{eHF}
\eea
for species $\sigma \neq \sigma'$, with partial densities
\be
 \rho_{\sigma}(r) =\sum_{nl}\,(2 l+1)\,{\,\vert u_{nl}^{\sigma}(r)\vert^2\over
4\pi r^2}\,
 n(\varepsilon_{nl}^{\sigma})
\ee
and trapping potentials $m\, \omega_{\sigma}^2\,r^2/2$.

For vanishing temperature, the Fermi-Dirac occupation numbers are step
functions limiting the summation to states below the respective Fermi sea
$\varepsilon_F^{\sigma}$ that fulfills the number equation $N_{\sigma} =
\sum_{\nu} \Theta(\varepsilon_F^{\sigma}-\varepsilon_{\nu}^{\sigma})$. In the
forthcoming calculations, we shall consider both equal as well as different
trapping frequencies $\omega_{\sigma}$ for species 1 and 2, 
the latter case devised to take into account the corresponding magnetic
projections of the trapped atoms, i.e., $(\omega_1/\omega_2)^2=
\sigma_1/\sigma_2$.\cite{trapp1}

\section{A simplified model for zero sound modes}
\label{sec:model_zero}
	
	In order to get  a basic understanding of the different
excitations of the two-component gas, we propose a very simple model for the
noninteracting  propagator that makes room to illustrative
analytical results. Let us assume that each of the
coexisting species possesses only one elementary excitation with energy
$\omega_{\sigma}$ (resp. $\omega_{\sigma'}$).  The corresponding free
propagator reads
\be
G_0^{\sigma \sigma}({\bf r}, {\bf r}') = F^{\sigma}({\bf r})\,\left[F^{\sigma}({\bf 
r}')\right]^*
\,\chi_0^{\sigma}
\label{G0trucho}
\ee
with
\be
\chi_0^{\sigma}=\frac{1}{\omega -\omega_{\sigma} + i \eta}
-\frac{1}{\omega+\omega_{\sigma}+i \eta}
\ee
being $F^{\sigma}({\bf r})$ the wave function of the excited ph pair at the
given position. Replacement of (\ref{G0trucho}) into the RPA system of
equations (\ref{sistema}) brings into evidence that these propagators are of
the form
\bea
G^{\sigma \sigma^{-1}}({\bf r}, {\bf r}') &=& 
F^{\sigma}({\bf r})\,\chi_0^{\sigma}\,\left[\Gamma^{\sigma \sigma^{-1}}({\bf 
r}')\right]^* 
\nonumber \\
G^{\sigma' \sigma^{-1}}({\bf r}, {\bf r}')
&=& g F^{\sigma'}({\bf r})\,\chi_0^{\sigma'}\,\left[\Gamma^{\sigma'  
\sigma^{-1}}({\bf r}')\right]^*
\label{sist2}
\eea
where in turn
\bea
\Gamma^{\sigma \sigma^{-1}} &=& 
F^{\sigma \sigma^{-1}} 
+ \tilde{g}^*\,\chi_0^{\sigma'}\,
\Gamma^{\sigma' \sigma^{-1}}
\nonumber\\
\Gamma^{\sigma' \sigma^{-1}}
&=& \tilde{g}^*\,\chi_0^{\sigma}\,
\Gamma^{\sigma \sigma^{-1}}
\label{sist3}
\eea
represent dressed ph wave functions, which depend on the ph interaction strength
\bea
\tilde{g} &=& g\,
 \int d{\bf x} \left[F^{\sigma}( {\bf x})\right]^*\,F^{\sigma'}({\bf x})
\nonumber
\\
&\equiv& \langle \sigma \sigma^{-1} \vert V_{\rm ph} \vert \sigma' \sigma'^{-1} 
\rangle
\label{gmonio}
\eea

 The system (\ref{sist3}) possesses a simple algebraic solution, from which we 
can write the transition densities as 
\bea
\delta \rho^{\sigma \sigma^{-1}}({\bf r}) &=& \frac {\chi_0^{\sigma}\,
 \left[F^{\sigma}({\bf r})\right]^* \,\theta^{\sigma}}
{D}
\nonumber \\
\delta \rho^{\sigma' \sigma^{-1}}({\bf r}) &= 
&\tilde{g}\,\frac{\chi_0^{\sigma}\,
\chi_0^{\sigma'}\,\left[F^{\sigma}({\bf r)}\right]^*\,\theta^{\sigma'}}{D}
\label{dro-model}
\eea
being
\be
D(\omega)= 1 - \vert \tilde{g}\vert^2\,\chi_0^{\sigma}(\omega)
\,\chi_0^{\sigma'}(\omega)
\label{D}
\ee
with $\theta^{\sigma} = \int d {\bf r}_i\,O^{\dagger}_i\,F^{\sigma}({\bf r}_i)$ the 
matrix element of the transition operator. Consequently, the dynamical 
susceptibilities are
\bea
\chi^{\sigma \sigma^{-1}}& = &\frac {\chi_0^{\sigma}\,
 \vert \theta^{\sigma}\vert^2}
{D}
\nonumber \\
\chi^{\sigma' \sigma^{-1}} &= &\tilde{g}\,\frac{\chi_0^{\sigma}\,
\chi_0^{\sigma'}\,\left[\theta^{\sigma}\right]^*\,\theta^{\sigma'}}{D}
\label{chi-model}
\eea

The appearance of the common denominator $D(\omega)$ in
Eqs.~(\ref{chi-model}) reflects the fact that the RPA  describes
collective fluctuations of the system as a whole. The zeroes of $D$ in 
(\ref{D}) can be easily found to be
\be
\omega_0^2=\frac{\omega_{\sigma}^2+\omega_{\sigma'}^2}{2}
\pm \vert \frac{\omega_{\sigma}^2-\omega_{\sigma'}^2}{2}\vert\,
\left[1+\frac{\displaystyle 16 \vert \tilde{g} 
\vert^2\,\omega_{\sigma}\,\omega_{\sigma'}}{\displaystyle 
\left(\omega_{\sigma}^2-\omega_{\sigma'}^2 \right)^2}
\right]^{1/2}
\label{omega0}
\ee
The evolution of the collective modes with interaction strength is
 encompassed in Eq. (\ref{omega0}); assuming for instance $\omega_{\sigma}^2 
\ge \omega_{\sigma'}^2$, to lowest order in the expansion parameter one finds,
\bea
\omega_{>}^2 &= &\omega_{\sigma}^2 + \frac{ 4 \vert \tilde{g} 
\vert^2\,\omega_{\sigma}\,\omega_{\sigma'}}
{\omega_{\sigma}^2-\omega_{\sigma'}^2}\,\,\ge \omega_{\sigma}^2
\nonumber \\
\omega_{<}^2 &= &\omega_{\sigma}^2 - \frac{ 4 \vert \tilde{g} 
\vert^2\,\omega_{\sigma}\,\omega_{\sigma'}}
{\omega_{\sigma}^2-\omega_{\sigma'}^2}\,\,\le \omega_{\sigma'}^2
\label{modos}
\eea
showing that within the validity of the approximation involved, the collective 
modes progress towards opposite directions with increasing  coupling. For large 
interaction strength and/or close intrinsic frequencies of the species, one can 
derive a complementary limit (cf. Eq. (\ref{omega0})) which gives
\be
\omega_0^2=\frac{\omega_{\sigma}^2+\omega_{\sigma'}^2}{2}
\pm 2 \vert \tilde{g} \vert \sqrt {\omega_{\sigma}\,\omega_{\sigma'}} 
\,\left[ 1 + \frac{\left(\omega_{\sigma}^2-\omega_{\sigma'}^2\right)^2}{32
 \vert \tilde{g}\vert^2\,\omega_{\sigma}\,\omega_{\sigma'}}\right]
\label{dos}
\ee

	The particular situation where the two interacting species are 
identified by the same well parameters is especially enlightening. On the one 
hand, one can see from (\ref{dos})  that if $\omega_{\sigma} = 
\omega_{\sigma'}$, the dispersion relation of the modes is simply given by

\be
\omega_0^2 = \omega_{\sigma}^2 \left( 1 \pm \frac{2 \vert \tilde{g}\vert 
}{\omega_{\sigma}}
\right)
\label{o}
\ee
which for sufficiently small relative strength  $\vert 
\tilde{g}\vert/\omega_{\sigma}$ illustrates the fragmentation of the elementary 
excitation into two collective states with energies
\be
\omega_0 = \omega_{\sigma} \pm \vert \tilde{g}\vert
\label{polo1}
\ee
The corresponding form of the total spin symmetric and antisymmetric 
responses  is (cf. Eq.~(\ref{chi_s_a}))
\be
\chi^{(s,a)} = 
\frac14\sum_{\sigma=\sigma_1,\sigma_2}\frac{\chi_0^{\sigma}\,\vert 
\theta^{\sigma}\vert^2}
{2\,\left(1 \pm \tilde{g}\,\chi_0^{\sigma}\right)}
\label{chi_s_a_zero}
\ee
(for real $\tilde{g}$). It is important to remark that  although both
frequencies (\ref{polo1}) correspond to poles of {\em each} term in
(\ref{chi_s_a_zero}), explicit computation shows that
$\omega_\sigma+|\tilde{g}|$ has zero amplitude in the antisymmetric response
while $\omega_\sigma-|\tilde{g}|$ has zero amplitude in the symmetric one.
The higher and lower frequencies in (\ref{polo1}) can be then attributed
to the symmetric and antisymmetric fluctuations respectively.
Their intensities, at the corresponding energies, can be derived as

\bea
S_0^{(s,a)} &=&\frac14\sum_{\sigma=\sigma_1,\sigma_2}\frac{\vert 
\theta^{\sigma} \vert^2\,\omega_{\sigma}}{4 
\omega_0}\,
\left(1 \pm \tilde{g}\,\chi_0^{\sigma}(\omega_0^{(s,a)})\right)
\nonumber
\\
&=&\frac14 \sum_{\sigma=\sigma_1,\sigma_2}\frac{\vert \theta^{\sigma} 
\vert^2\,\omega_{\sigma}}{2 \omega_0}
\label{S0}
\eea
The latter line holds in view of (\ref{o}) and coincides with the  unperturbed 
strength of the elementary excitation. 

In summary, this simple model gives rise to spectrum fragmentation into two
lines, according to Eq.~(\ref{omega0}); in the specific case of equal trapping
frequencies and in the weak coupling limit, we see that due to amplitude
cancellation, one can clearly identify {\em different} excitations energies in
the $s$ and $a$ channels.

\section{Calculations and results}
\label{sec:calcula}

	We have solved the HF problem for variable numbers of $^{40}$K atoms
$N_1, N_2$ corresponding to different spin projections in an isotropic
harmonic trap, with a mutual $s$-wave scattering length $a$ = 8.31 nm
\cite{pwave}.  As in Ref.  \cite{bruun1}, we start an iterative procedure from
oscillator wave functions $u_{nl}^{osc}(r)$, and convergence is rapidly
achieved. The selfconsistent states are labelled by the same quantum numbers
$nl$ and the wave functions differ only slightly from the original ones. The
low energy states are the most sensitive to the size of the interaction
strength, reaching deviations with respect to bare oscillator energies as
large as 15\%.  The combinations of ph states entering the free ph propagator
(see Eq.~(\ref{G0})) are selected by angular momentum conservation,  and we
find that the elementary excitation energies in the denominators of the
generalized susceptibilities (see Eq.~(\ref{chinunu'})) are weakly spread
around the noninteracting oscillator values.

\subsection{Equal trapping potentials and populations}

As a first step we examine a trapped two-component Fermi gas with equal trapping
potentials (ETP), i.e. $\omega_1=\omega_2$, and the same number of atoms in each
hyperfine level. Under these conditions the RPA equations
(\ref{sistema}) can be decoupled for $\delta\rho_{\sigma}^{(s)}$ and
$\delta\rho_{\sigma}^{(a)}$, giving rise to

\bea
\delta\rho^{(s,a)}(r)&=&\delta\rho^{(s,a)}_{0}(r)
\nonumber
\\
& \pm&
\frac{4\pi}{2L+1} g \int\!\! r'^2\,G_{0\,L}(r,r',\Omega)\,
\delta\rho^{(s,a)}(r') 
\label{desacopla}
\eea
where we identify $\delta\rho^{(s,a)} \equiv \delta\rho_{\sigma}^{(s,a)}$ and
$G_{0\,L} \equiv G^{\sigma\,\sigma^{-1}}_{0\,L}$ for either spin projection 
$\sigma$. Equations
(\ref{desacopla}) are identical to the RPA equations for a {\em single}
component gas interacting through a repulsive (resp. attractive)
contact interaction, giving rise to the symmetric (resp. antisymmetric) fluctuation. 

\begin{figure}
\centering
\epsfig{file=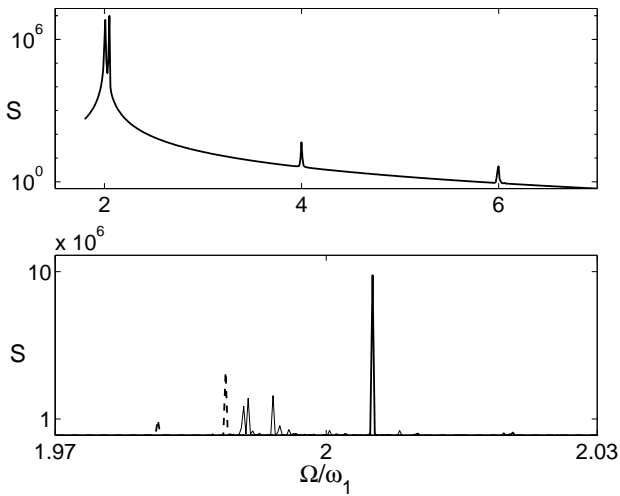,width=.95\columnwidth,clip=}
\caption{Monopolar dynamic structure factor (in arbitrary 
units) in the ETP case for $N_1=10^4$ and $\omega_1=2\pi \times 70\,s^{-1}$.
The upper plot is depicted in logarithmic scale and the lower one is a zoom of
the main peak in linear scale. The thin line corresponds to the HF 
ph excitations, and thick and dashed lines indicate the RPA structure factor in 
the symmetric and antisymmetric channels, respectively. } 
\label{fig:eq1} 
\end{figure}

 In Fig.~\ref{fig:eq1} we show the strengths $S^{(s,a)}$ of the monopolar
($L=0$) fluctuations for $N_{1,2}=10^4$. It should be kept in mind that the
spread of the peaks in $S^{(s, a)}(\Omega)$ is an artifact of the calculation,
due to the introduction of the small numerical parameter $\eta$ (cf.
Eq.~(\ref{chinunu'})).
The large scale behavior of $S(\Omega)$ shows multiple excitations at
energies close to $\Omega_n=\varepsilon_{n_1\,l}-
\varepsilon_{n_2\,l}\approx2\,n\,\omega_1$ . These peaks decrease their
amplitude (notice the logarithmic scale) as we increase the transferred
energy. Although the $s$ and $a$ channels are different, we are not able to
visualize them in the current scale. A narrow region around the oscillator
excitation energy  is displayed in the lower plot of Fig.~\ref{fig:eq1}. In
fact, a careful analysis of the peaks indicates that they group into two main
sets. Only the strongly fragmented one around $\Omega=2\omega_1$ can be
viewed in this figure; the intensities in the second group lying at higher
energies are too small in the current scale.  Moreover, the spectra in both
channels are clearly different: while the $s$ channel is shifted upwards in
energy, the $a$ one lies at lower energies, in qualitative agreement with the
simple model estimate of the preceding section.  It is important to mention that
the qp's energies coming out from the HF calculation are such  that the ph
excitation frequencies are lowered with respect to the oscillator ones by the
mean field interaction, while the ph coupling introduced in the RPA formalism
shifts the collective spectrum to higher energies. In addition, we have
numerically verified that the position of the main peak can be accounted for by
the analytical sum rule formula derived by Vichi and Stringari \cite{vichi3}. 

\begin{figure}
\centering
\epsfig{file=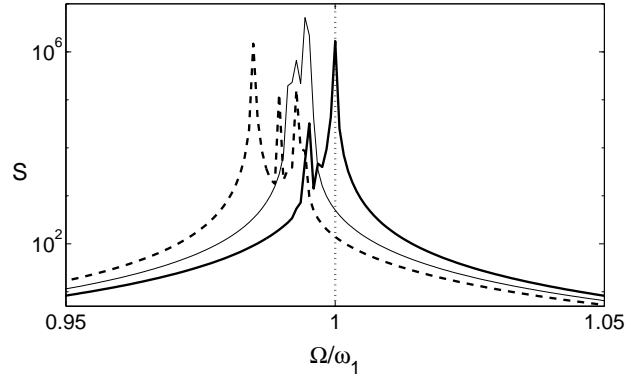,width=0.95\columnwidth,clip=}
\caption{Dipolar dynamic structure factor (in logarithmic scale and arbitrary
units) for the same conditions as in Fig.\protect\ref{fig:eq1}. Thin, thick
and dashed lines respectively correspond to the HF system, symmetric and
antisymmetric channels.}
\label{fig:eq3}
\end{figure}

The behavior of the dipolar fluctuation ($L=1$) is slightly different.
Although both the $s$ and $a$  spectra are weakly fragmented, the amplitude of
the oscillator mode at $\Omega =\omega_1$ is large. It is well known
\cite{gen} that for equal trapping potentials and populations there is a
dipolar excitation in the symmetric channel, associated to the center-of-mass
oscillation of the gas and occurring at the oscillator frequency. However, in
the spin-like channel this excitation appears at a slightly lower energy.
These facts are verified in the present RPA calculation as shown in
Fig.~\ref{fig:eq3}, where we see the dipolar structure factor for
$N_1=N_2=10^4$ atoms. We can also observe  that in addition to these modes,
fragmented poles of collective nature show up as indicated  by the simple
model of the preceding section.   

\begin{figure} 
\centering 
\epsfig{file=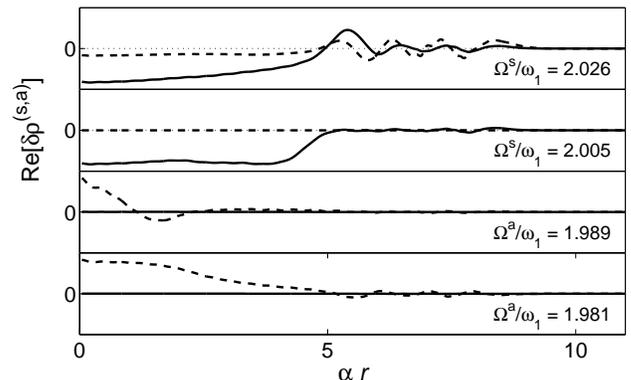,width=.95\columnwidth,clip=}
\caption{Monopolar density fluctuations for equal trapping potentials 
and populations (in arbitrary units) with $N_{1,2}=10^4$.  Solid and 
dashed lines respectively correspond to density and spin-like excitations 
at the poles of each channel. $\alpha^{-1}$ is the distance unit for 
the trapping potential of species 1, $\alpha=(\hbar/m\omega_1)\approx 0.53 
\mu{\rm m}^{-1}$.} 
\label{fig:drosa}
\end{figure}

Let us now examine the spatial profiles of the density fluctuations. In
principle, one would expect that if the system is excited with a given
operator $s$ or $a$ (cf. Eq.~(\ref{eq:operators})) in the vicinity of a peak
in one of these channels, a density fluctuation will develop that reflects
both the character of the external field and the nature of the intrinsic
excitation of the free system. As an illustration, in Fig.~\ref{fig:drosa} we
show the real parts of the monopolar density oscillations for the same
conditions as in Fig.~\ref{fig:eq1} at the frequencies given by the poles of
the response. We observe that when the symmetric fluctuation is important
within the bulk of the trapped system, the antisymmetric counterpart is
completely negligible and viceversa. This is in agreement with the
interpretation of the simplified model of Sec.~\ref{sec:model_zero} in which
density and spin-like fluctuations were clearly distinguished. However, at the
energy of the pole in the symmetric channel ($\Omega^s \approx 2.03$), surface
oscillations develop both in $\delta\rho^s$ and $\delta\rho^a$; this behavior
can be attributed to the proximity of a weak peak in the antisymmetric
structure factor (not visible in the scale of Fig.~1).

\subsection{Unequal potential wells and populations}

In the recent experimental achievement of DeMarco and Jin \cite{demarco},
$^{40}$K atoms in two magnetic sublevels $|F=9/2, m_F=7/2\rangle$  (type 1
atoms), and $|F=9/2, m_F=9/2\rangle$ (type 2 atoms), were confined and cooled.
Although the ratio $\omega_2/\omega_1=(\sigma_2/\sigma_1)^{1/2}\approx 1.13$
is very close to unity, we take it into account as an explicit feature of the
real nonsymmetric configuration. We also consider unequal populations ranging
from $\Delta=N_2/N_1\approx 0.3$ to $\Delta\approx 3$,  which can be built
during the evaporative cooling process. 

\begin{figure}
\centering
\epsfig{file=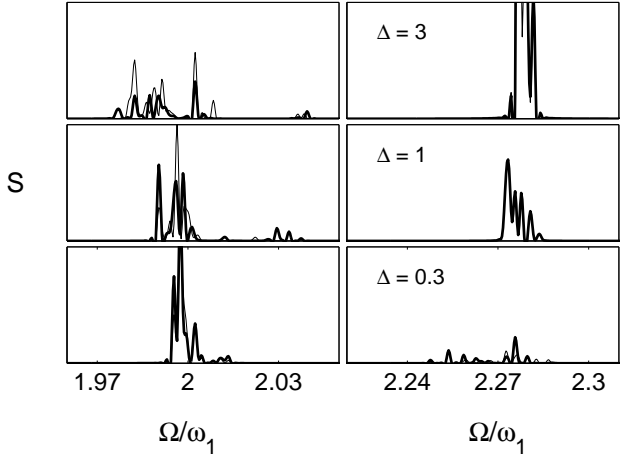,width=.95\columnwidth,clip=}
\caption{Dynamic structure factor for $L=0$ (in arbitrary units) for the 
symmetric channel of the interacting and free system with $N_1=10^4$ and 
several concentrations, in thick and thin  
lines, respectively. Each column display a different range in the energy
scale.}
\label{fig:S0}
\end{figure}

As in the ETP case we analyze the dynamic structure factors and the
fluctuations created by multipolar external fields. In Fig.~\ref{fig:S0} we
show the symmetric dynamic structure factor for several values of the
population ratio $\Delta$, $N_1=10^4$ and a monopolar excitation.  The
structure of spin-like fluctuations cannot be distinguished from the
density-like ones within the scale of this plot.  The essential features of
the response can be summarized as follows: i) in either channel, the strength
of each ph transition is redistributed so that the structure factor is
fragmented around the noninteracting peaks with some intensity appearing at
higher energies; ii) for low or high values of  $\Delta$ the structure of the
most populated species essentially reproduces the pattern of the
noninteracting quasiparticles, while the spectrum of the other species is
highly fragmented with a largely suppressed amplitude.  iii)  given either
species, as the number of atoms in the other spin projection increases, the
complementary excitation appears displaying considerable fragmentation. 

\begin{figure}
\centering
\epsfig{file=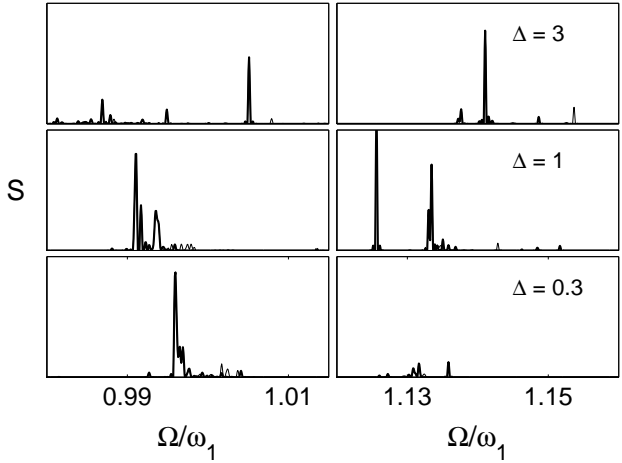,width=.95\columnwidth,clip=}
\caption{Same as Fig.\protect\ref{fig:S0} for the dipolar dynamic structure factor.}
\label{fig:S1}
\end{figure}

\begin{figure}
\centering
\epsfig{file=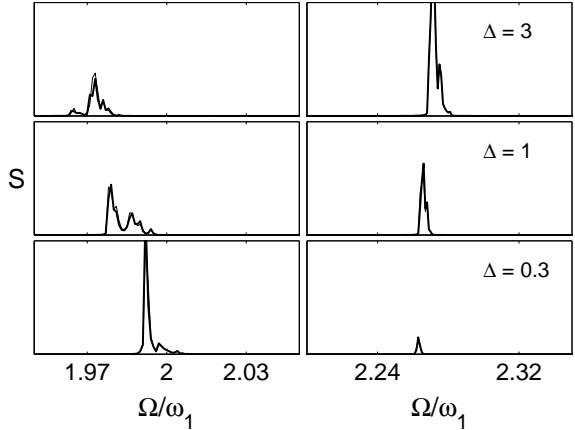,width=0.95\columnwidth,clip=}
\caption{Same as Fig.\protect\ref{fig:S0} for the quadrupolar  dynamic structure factor.}  
\label{fig:S2}
\end{figure}

 We have also made calculations for multipolar excitations with $L=1,2$ ; in
Figs.~\ref{fig:S1} and ~\ref{fig:S2} we show the corresponding results for the
structure factor.  In general, the description of the monopolar excitation
applies as well to higher multipolarities, however we can mention some
differences: the dipolar peaks are narrower and fragmentation seems to be
stronger. In particular, in the equal population case ($\Delta=1$) there exist
well resolved fragmented peaks as intense as the original HF ones.

\begin{figure}
\centering
\epsfig{file=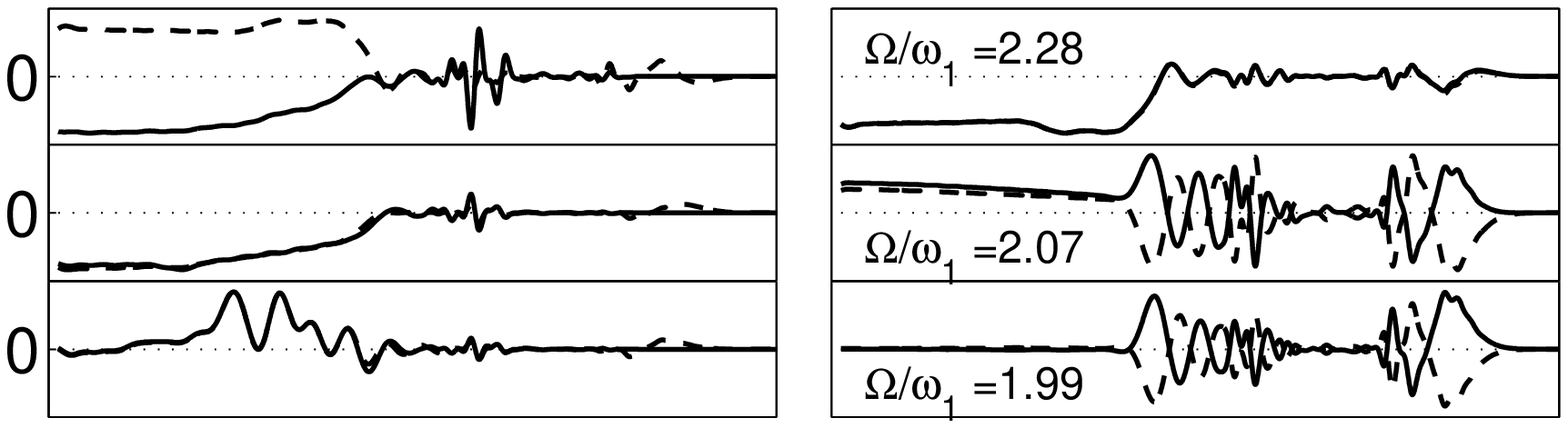,width=.95\columnwidth,clip=}
\epsfig{file=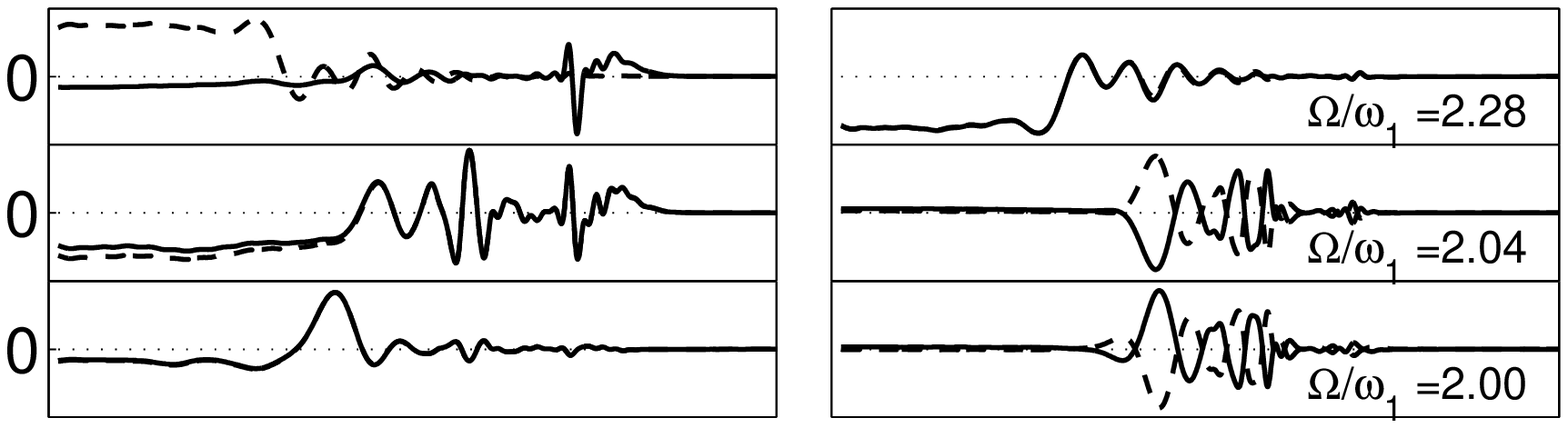,width=.95\columnwidth,clip=}
\epsfig{file=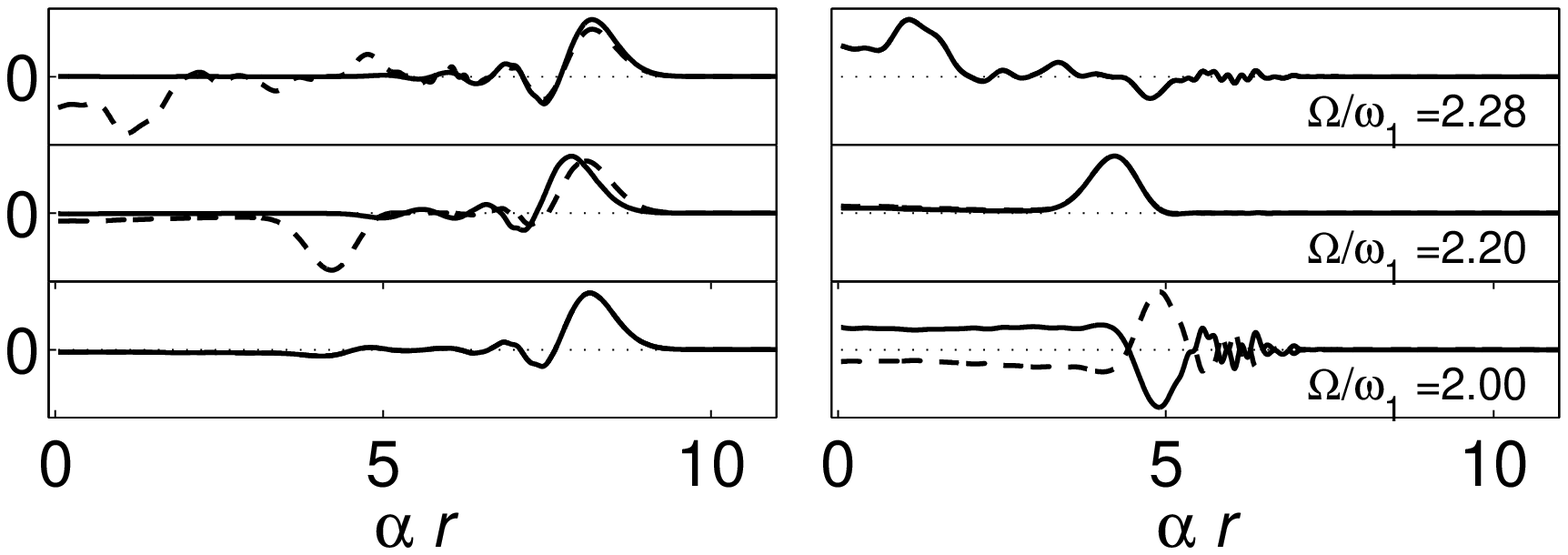,width=.95\columnwidth,clip=}
\caption{Monopolar density fluctuations (in arbitrary units) for $N_1=10^4$.
 The upper, middle, and  lower plots correspond to
$\Delta = 3, 1, 0.3$ respectively, for $\sigma_1 = 7/2$ (left column)
and $\sigma_2=9/2$ (right column). Solid and dashed  lines respectively indicate
$\delta\rho_{\sigma}^{(s)}$ and $\delta\rho_{\sigma}^{(a)}$.}
\label{fig:dro0}
\end{figure}

\begin{figure}
\centering
\epsfig{file=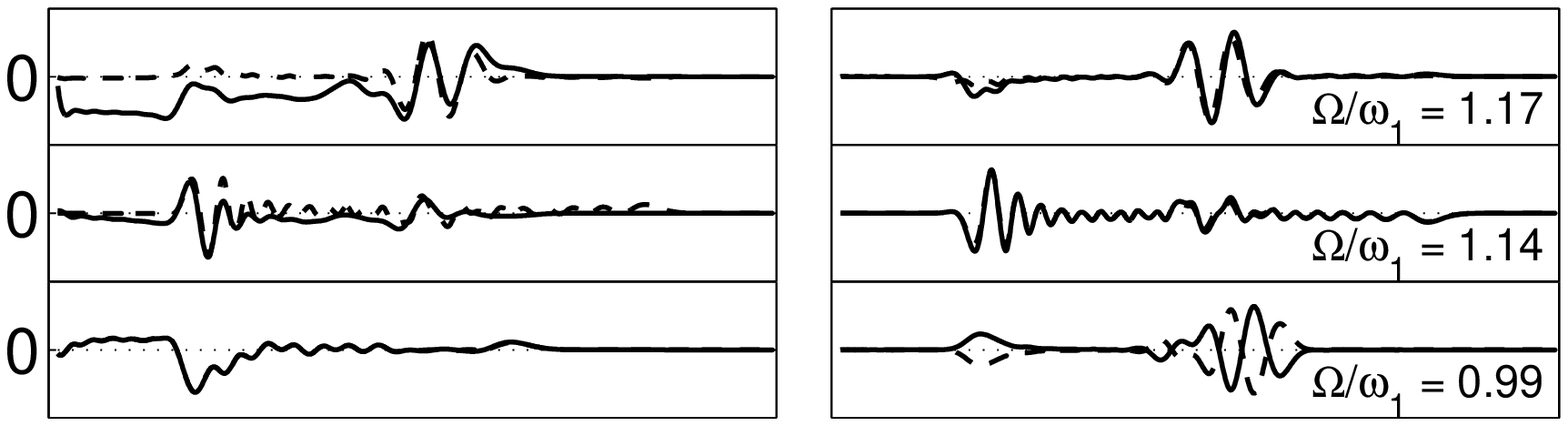,width=0.95\columnwidth,clip=}
\epsfig{file=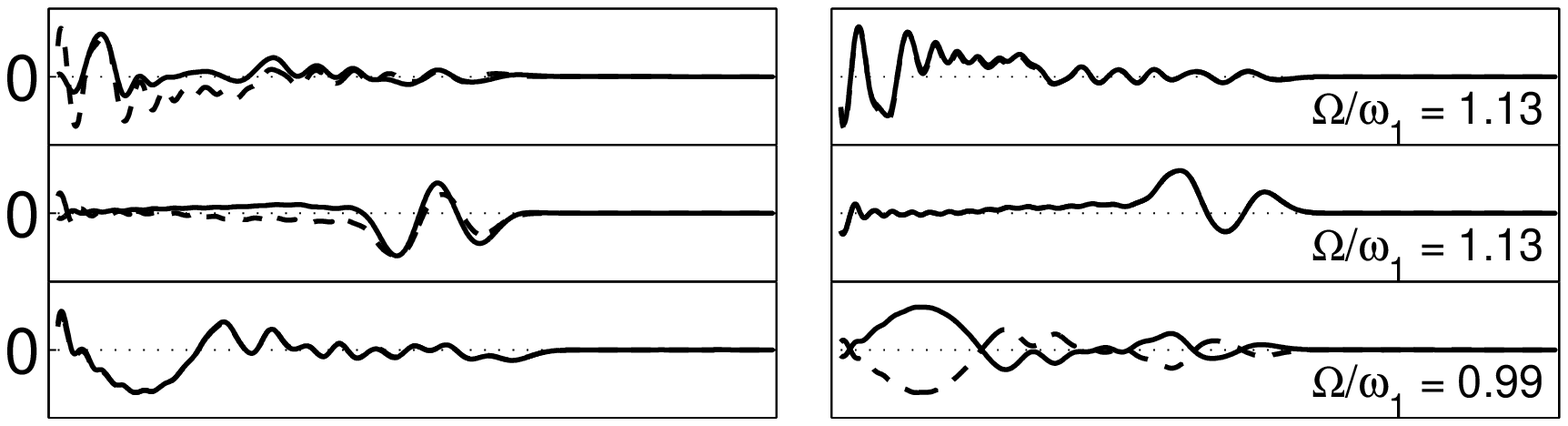,width=0.95\columnwidth,clip=}
\epsfig{file=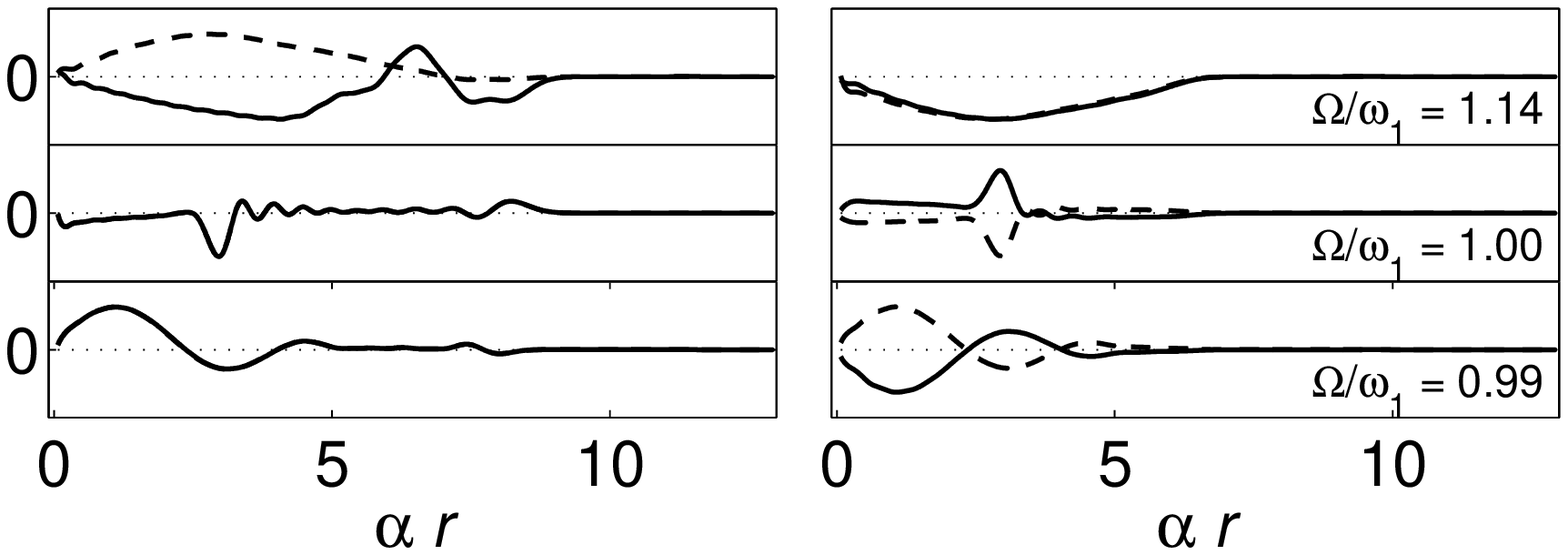,width=0.95\columnwidth,clip=}

\caption{Same as Fig.~\ref{fig:dro0} for dipolar excitations.} 
\label{fig:dro1}
\end{figure}

\begin{figure}
\centering
\epsfig{file=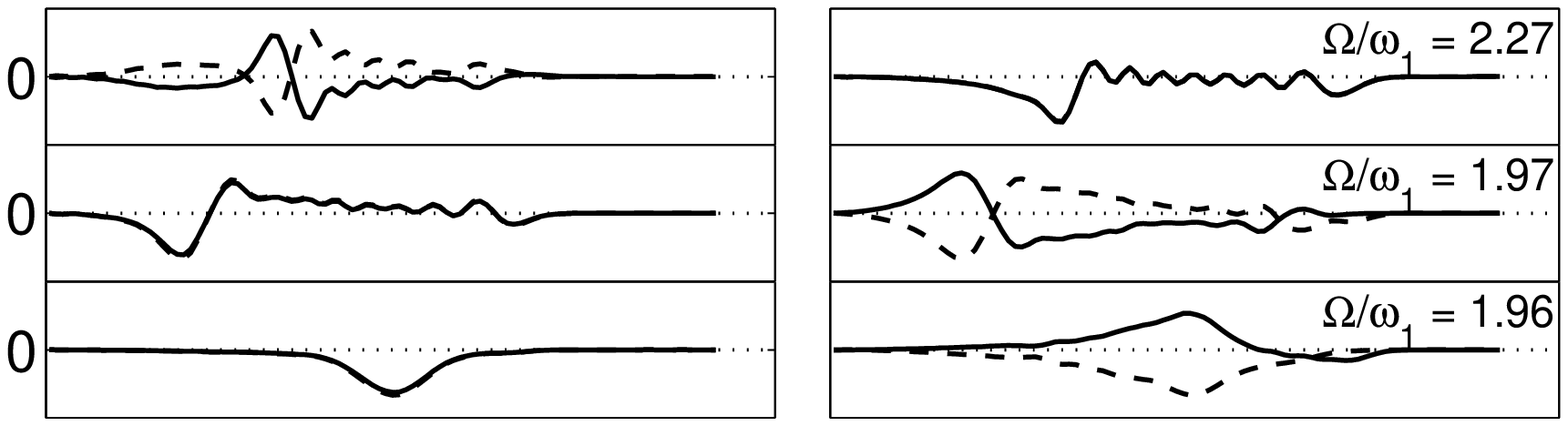,width=0.95\columnwidth,clip=}
\epsfig{file=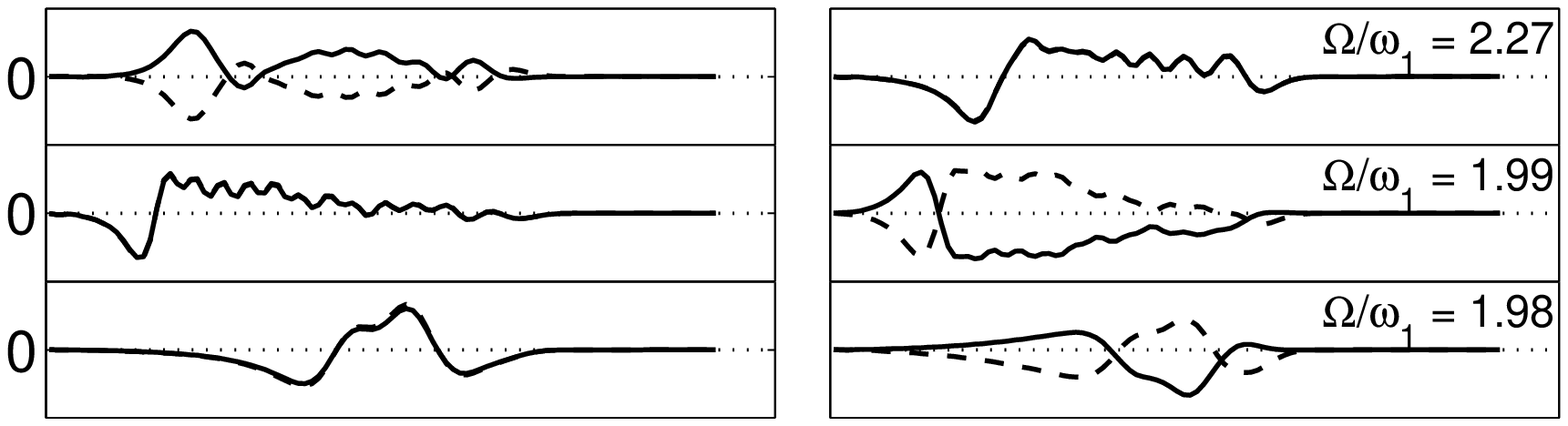,width=0.95\columnwidth,clip=}
\epsfig{file=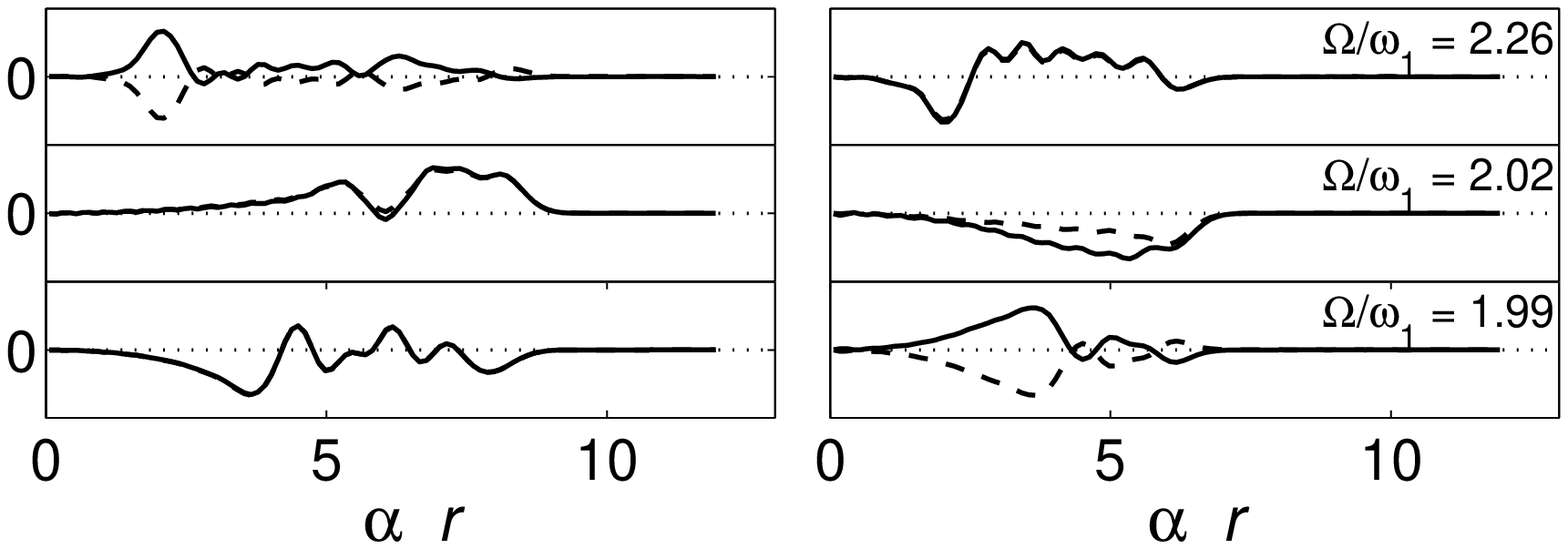,width=0.95\columnwidth,clip=}
\caption{Same as Fig.~\ref{fig:dro0} for quadrupolar  excitations.}
\label{fig:dro2}
\end{figure}

The real parts of the transition densities are displayed in
Figs.~\ref{fig:dro0} to ~\ref{fig:dro2} for $L$ = 0 to 2, respectively. The  
$s$ and $a$ density profiles for each component have been scaled to a 
common value at a given frequency. This is a convenient criterion, since if one depicts the four
profiles in the same scale, near the intrinsic ph frequencies of each component
the complementary density fluctuation appears largely depleted. 
Before analyzing the fluctuation profiles, we want to bring some attention to
the shape of the  HF equilibrium density profiles. In Table~\ref{tab:1} we
quote, as a function of $\Delta$, some shape parameters related to the
spherical probability density distribution $P_\sigma(r)$ defined as:
\be
P_\sigma(r) = {r^2\,\rho_\sigma(r)\over N_\sigma}
\label{pdf}.
\ee
Particularly, we list the maximum probability $P_{\rm max}$ which is attained 
at $R_{\rm max}$, the location of the density edge $R_{\rm edge}$ and the 
full width at half maximum (FWHM) of the probability distribution. These 
parameters will help us to understand the main aspects of the excitation 
profiles.

In Fig.~\ref{fig:dro0} we show some typical monopolar density fluctuations for
each spin component, at the lowest-lying collective peaks and several
concentrations. We observe that as we increase $\Delta$ (from bottom to top in
Fig.~\ref{fig:dro0}), both fluctuations $\delta\rho_2^{(s,a)}$ for a given frequency
extend beyond the $\rho_2$ edge.  As an illustration, let us consider in
detail the case $\Delta=0.3$. For  the excitation with
$\Omega\approx 2\,\omega_1$, $\delta\rho_2^{(s, a)}$ is bounded
to roughly the size of the density of $\sigma_2$ atoms ($\alpha r \lesssim 7$,
$\alpha^2=\hbar/m\omega_1$ ) while $\delta\rho_1^{(s, a)}$ extends far beyond
this cutoff, within the $\rho_1$ range.  For small $\Delta$ the corrections
$\delta\rho^{1\,2^{-1}}$ are unimportant, in general, due to the relatively
weak coupling of the ${\sigma}_1$ species to the few $\sigma_2$ atoms.
However, in the same limit $\delta\rho^{2\,2^{-1}}$ acquires important
corrections $\delta \rho^{2\,1^{-1}}$,  essentially driven by the large number
of $\sigma_1$ atoms.  We also see two different behaviors related to the
$\Omega$ dependence: 
for frequencies close to 2$\omega_1$, $\delta\rho_2^{(s)}$
and $\delta \rho_2^{(a)}$ present opposite signs, revealing that the induced  
contribution $\delta\rho^{2\,1^{-1}}$ is larger than the intrinsic one
$\delta\rho^{2\,2^{-1}}$.  In fact, if the system is excited with frequencies
close to $2\,\omega_1$, large $\delta\rho_1$ fluctuations should be expected,
which in turn introduce, through  the $V_{\rm ph}$ interaction, sizeable
$\delta \rho^{2\,1^{-1}}$ contributions to  $\delta\rho_2^{(s, a)}$.  In turn,
exciting at $\Omega \approx 2\,\omega_2$ creates $\delta\rho^{2\,2^{-1}}$,
thus the correction to $\delta \rho^{2\,1^{-1}}$ is second order in the
interaction and the $s$ and $a$ fluctuations are the same. On the other hand,
as $\rho_1$ extends beyond $\rho_2$, a stimulus acting at
$\Omega=2\,\omega_2$ induces a cross-fluctuation $\delta\rho^{12^{-1}}$
inside the bulk of the type 1 system, i.e., at smaller radii (cf.
Fig.~\ref{fig:dro0}).

As  one increases the number of particles in species 2, their  density
profiles extend farther (see Table~\ref{tab:1}) and there are few noticeable
changes in the spatial localization of the excitation. However, in general,
close to $2\,\omega_2$ we find a strong induced fluctuation
$\delta\rho^{12^{-1}}$ in the spatial region where $P_2(r)$ is larger than
$P_1(r)$ and correspondingly for $\delta\rho^{21^{-1}}$ near $2\,\omega_1$.
If one keeps increasing $\Delta$, the perturbative interpretation
 is no longer valid for every multipole and species. For $\Delta=3$,
the fluctuations occupy a broader region, roughly from $\alpha r=4$ to
$\alpha r= 10$, this can be attributed to a wider $\rho_2$ density.   

In addition to the vanishing at origin of the $L\neq0$ fluctuation, the main
difference between the spatial profiles of distinct multipolarities, displayed
in Figs.\ \ref{fig:dro0} to \ref{fig:dro2}, lies in the enhanced symmetry
of the channels for $L=2$. In this case, we observe that the
cross-fluctuation of a given component is non-negligible {\em only} near the
free ph transition of its counterpart, suggesting a weaker ph effective
interaction for higher multipolarities (cf. Eq.\ (\ref{dro})).   

Another feature to remark is the difference with the ETP case; in that
situation we have the same spatial fluctuations for either spin projection and
 well defined channels $s$ and $a$; however, in this more
general problem, we observe different spatial profiles for each component and 
mixed behaviors within the channels, i.e., given a pole in the
symmetric susceptibility, fluctuations may exhibit similar amplitudes in
both channel.

\section{Discussion and summary} 
\label{sec:resumen}

In this work we have developed a RPA formalism for a two-species, trapped
Fermi gas at vanishing temperature,  which may provide guidelines to current
experimental research. We have shown that the interspecies interaction gives
rise to a fragmented  zero sound spectrum.  We analyzed both the equal
trapping potential and populations, as well as a general case with unequal
potentials and populations intended to mimic an experimental situation.  The
main differences arise in the spectra; while in the ETP case the poles
associated to density-like and spin-like fluctuations were clearly distinct,
in the general case these fluctuations cannot be disentangled.  The
density-like and spin-like responses of the system share the same energy
spectrum, however with unequal amplitudes.  We have presented a  simplified
picture for the ph propagator and elementary spectrum which gives support to
the numerical results. This model also allows us to interpret the frequency
shifts in Eq.~(\ref{polo1}) in terms of the interaction among pair
excitations, measured by the product  $\tilde{g}$ of the bare coupling
strength times the overlap between the pair wave-functions (cf.  Eq.
(\ref{gmonio})).  For the excitation with $L=1$ we have found in the density
channel and  for the ETP case a pole at the bare oscillator frequency
corresponding to rigid oscillation of the system; however this is no longer
true for  a more realistic configuration with unequal trapping frequencies.
Although we checked that the eigensolutions of Eq.~(\ref{eHF})  differ only in
a few percent from the bare oscillator basis functions, we cannot prevent
propagation of small errors in the response calculation, which ought to be
safely computed out of the true HF wave functions to the price of a increase
in computing time. Finally, we should mention that this two-component RPA 
formalism may be straightforwardly generalized to nonzero temperatures 
by taking into account the finite temperature Fermi-Dirac occupation numbers.

\acknowledgements

This paper was supported by Grant No. PICT 1706 from Agencia Nacional de
Promoci\'on para la Ciencia y la Tecnolog\'{\i}a of Argentina and Grant No.
TW81 from the Universidad of Buenos Aires (UBA). One of us (P.C.) is grateful
to the Universidad de Buenos Aires for financial support.

\appendix
\section*{}

To evaluate the free ph propagator we use the 3D harmonic oscillator basis 
for each spin projection:
\bea
\phi_{nlm}^{\sigma}({\bf r})&=&
A_{nl}^{\sigma}\,\exp(-\alpha_{\sigma}^2r^2/2)\,r^l\,L_n^{l+1/2}(\alpha_{\sigma}^2\,r^2)\,Y_{lm}
(\hat{r})\nonumber \\
&\equiv&R_{n\,l}(r)\,Y_{lm}(\hat{r})
\eea
being $\alpha_{\sigma}^2=\hbar/m \omega_\sigma$, $Y_{lm}$ spherical harmonic
functions, $L_n^{l+1/2}$ the generalized Laguerre functions and 
$A_{nl}^{\sigma}$ a normalization constant.  The
multipolar component $G_{0\,L}^{\sigma\,\sigma^{-1}}$ reads
\bea
&&G_{0\,L}^{\sigma\,\sigma^{-1}}(r,r',\Omega)=\frac1{(4\pi)^2}\,\sum_{n\,l,n'\,l'}
(2l+1)(2l'+1)\,
\nonumber \\
&&(R_{n\,l}R_{n'\,l'})(r)\,(R_{n\,l}R_{n'\,l'})(r')\,
\vert\langle
l\,0\,l'\,0\vert L\,0\rangle\vert^2\,\chi_{n\,l,n'\,l'}(\Omega)
\nonumber \\
\label{G0L}
\eea
In order to compute the elementary susceptibilities $\chi_{n\,l\;,n'\,l'}$ in
Eq.(\ref{G0L}), we have used the HF energies extracted from (\ref{eHF})
instead of the bare harmonic oscillator ones. Further simplifications 
arise from the properties of the ClebschGordan coefficients and HO
wave-functions $R_{nl}(r)$;  in particular, explicit expressions 
for monopolar, dipolar and quadropolar excitations ($L=0,1,2$) can be written. 
The monopolar free ph propagator reads  
\bea
G^{\sigma\,\sigma^{-1}}_{0\,0}(r,r',\Omega)&=&\frac1{(4\pi)^2}\,\sum_{n\,l\,n'}
\,(2 l +1)\,
R_{n\,l}(r)\,
R_{n'\,l}(r)\, \nonumber \\
& &R_{n\,l}(r')\,R_{n'\,l}(r)\,\chi_{n l,n' l}(\Omega)
\label{G0}
\eea
In this case we use the excitation operator $O_i^\dag=r_i^2$; this yields  
the following noninteracting susceptibility 
\bea
\chi_{0\,0}^{\sigma\,\sigma^{-1}}&=&\frac1{\alpha_\sigma^2}\sum_{n\,l}
(2 l+1)\left[
\chi_{n l , n-1\, l}\,n (n+l+1/2) \right. \nonumber \\
&+&\left. \chi_{n l , n+1 l}\,(n+1) (n+l+3/2)\right]
\eea
which in the case of large enough number of particles, zero temperature  and 
HO energies can be simplified to yield
\bea
S_{0\,0}^{\sigma}(\Omega)&=&-\frac1{\pi}\,{\rm  
Im}[\chi_0^{\sigma\,\sigma^{-1}}]
\nonumber \\
&\approx&
\frac{1}{\alpha_\sigma^2}\frac{3\,N_\sigma}{4}(6N_\sigma)^{1/3}\,\delta(\Omega-2\,w_{
\sigma}) 
\eea

Similarly, the dipolar propagator can be written as: 
\bea
G_{01}^{\sigma\,\sigma^{-1}}(\Omega)=\frac3{(4\pi)^2}\sum_{n\,l\,n'}
\left[(l+1)\, R_{n\,l}(r) R_{n'\,l+1}(r) \right. \nonumber  \\ 
\left. R_{n\,l}(r') R_{n\,l+1}\,\chi_{n l , n'\,l+1}(\Omega)
+ l\, R_{n\,l}(r)\right.\nonumber \\ 
 \left.  R_{n\,l-1}(r) R_{n\,l}(r') 
R_{n\,l-1}(r')\,\chi_{n l , n'\,l-1}(\Omega)\right]
\label{dipolar}
\eea
yielding a temperature independent free response 
\be
\chi_{0\,1}^{\sigma\,\sigma^{-1}}=\frac{1}{\alpha_{\sigma}^2}\frac{3N_\sigma}{8\pi}\left(
\frac1{\Omega-\omega_\sigma + {\rm i}\eta}
-\frac1{\Omega+\omega_\sigma +{\rm i}\eta}\right)
\ee
with a $T=0$ structure factor 
\be
S_{0\,1}^{\sigma}=\frac{1}{\alpha_\sigma^2}\frac{3N_\sigma}{8\pi}\delta\left(\Omega-\omega_\sigma\right)
\ee

\squeezetable
\begin{table}
\caption{Shape parameters related to the HF probability density profiles
$P_\sigma(r)$ (see text) for $N_1=10^4$  and several concentrations. All 
distances are given in harmonic oscillator units $\alpha^{-1}=(\hbar/m
\omega_1)^{-1}\approx 1.9\mu{\rm m}$.}
\label{tab:1}
\begin{tabular}{l|dddddd}
 & \multicolumn{2}{c}{$\Delta = 0.3$} & \multicolumn{2}{c}{$\Delta =
1$ }& \multicolumn{2}{c}{$\Delta = 3$} \\
Parameter& $\sigma_1$ & $\sigma_2$ & $\sigma_1$ & $\sigma_2$ & $\sigma_1$ & $\sigma_2$ \\
\hline
$P_{\rm max}$ & 0.017 & 0.022 & 0.017 & 0.018 & 0.017 &  0.015 \\
$\alpha R_{\rm max}$ & 5.64 & 4.28 & 5.64 & 5.24 & 5.64 & 6.28 \\
$\alpha R_{\rm edge}$ & 8.71 & 6.59 & 8.75 & 8.17 & 8.79 & 9.86 \\
FWHM & 5.04 & 3.92 & 5.04 & 4.72 &4.96 & 5.68 
\end{tabular}
\end{table}

\end{document}